\documentclass[12pt]{article}
\usepackage{geometry}                
\usepackage{graphicx}
\usepackage{amssymb}
\usepackage{cite}
\usepackage{amsmath}

\def\lsim{\mathrel{\rlap{\lower3pt\hbox{\hskip0pt$\sim$}}
    \raise1pt\hbox{$<$}}}
\def\gsim{\mathrel{\rlap{\lower4pt\hbox{\hskip1pt$\sim$}}
    \raise1pt\hbox{$>$}}}
\begin{document}
\newcommand{\beq}{\begin{equation}}
\newcommand{\eeq}{\end{equation}}
\def\beqn{\begin{eqnarray}}
\def\eeqn{\end{eqnarray}}
\newcommand{\smallfrac}[2] {\mbox{$\frac{#1}{#2}$}}
\def\slsh#1{\setbox0=\hbox{$#1$}%
\rlap{\ifdim\wd0>.7em\kern.22\wd0\else\kern.1\wd0\fi /}#1}
\newcommand{\qt}{\tilde q}
\newcommand{\E}{{\cal E}}
\newcommand{\qtu}{\tilde q_{1}}
\newcommand{\qtd}{\tilde q_{2}}
\newcommand{\ntwo}{${\cal N}=2\;$}
\newcommand{\none}{${\cal N}=1\;$}
\newcommand{\vp}{\varphi}
\renewcommand{\theequation}{\thesection.\arabic{equation}}


\begin{titlepage}
\renewcommand{\thefootnote}{\fnsymbol{footnote}}

\begin{flushright}
FTPI-MINN-06/01\\ 
UMN-TH-2426/06\\
IPPP/06/03\\ 
DCPT/06/06\\
hep-th/0602004
\end{flushright}


\vfil

\begin{center}
\baselineskip20pt
{ \Large \bf  Central Charge Anomalies in 2D Sigma Models with
Twisted Mass}
\end{center}
\vfil
\begin{center}


{ \large {M.~Shifman,$^{a}$ A. Vainshtein}},$^{a}$ and { \large {R.~Zwicky}}\,$^{b}$

\vspace{0.3cm}

{\it  $^{a}$\,William I. Fine Theoretical Physics Institute,
University of Minnesota,
Minneapolis, MN 55455, USA\\
$^{b}$\,IPPP, Department of Physics, University of Durham,\\ Durham DH1 3LE, UK
}

\vfil

\end{center}

\vspace*{.25cm}
\vfil
\begin{abstract}
We discuss the central charge in supersymmetric  ${\cal N}=2$ sigma models  in two  dimensions.
The target space is a symmetric  K\"ahler manifold; CP$(N\!-\!1)$ is an example. 
The  U(1) isometries allow one to introduce twisted masses in the model.
At the classical level the central charge  contains Noether charges  of the U(1) isometries
and a topological charge which is an integral of a 
total derivative of the Killing potentials.
At the quantum level the topological part of the central charge acquires anomalous terms. 
A bifermion term was found previously, using supersymmetry which relates it to 
the superconformal anomaly. We present a direct calculation of this term
using a number of regularizations. 
We derive, for the first time, the bosonic part in the central charge anomaly.
We construct the supermultiplet of all anomalies and present its superfield description.
We also discuss a related issue of  BPS solitons in the CP(1) model and present 
an explicit form for the curve of marginal stability.

\end{abstract}

\end{titlepage}

\section {Introduction}
\label{intro}
\setcounter{equation}{0}

It is well known that supersymmetric theories may have
BPS sectors in which some data can be computed 
at strong coupling even when the full theory is not solvable. Historically, 
this is how the first exact results on particle spectra  were obtained
\cite{A4}. Seiberg--Witten's breakthrough 
results \cite{SW1,SW2} in the mid-1990's gave an additional 
motivation to the studies of the BPS sectors.

BPS solitons can emerge in those supersymmetric theories
in which superalgebras are centrally extended.
In many instances the corresponding central charges are seen at
the classical level.  In some interesting models central charges
appear as quantum anomalies.  Witten suggested in 1978 that such 
central charge should arise in two-dimensional CP$(N\!-\!1)$ models \cite{Wcp}. 
His conjecture was based on the fact that
the solution he obtained in the $1/N$ expansion revealed 
the BPS nature of the soliton supermultiplets.
Rather recently  \cite{LS-03} the central charge 
responsible for the multiplet shortening was identified as 
$\int dz (\partial/\partial z) \, (R_{i\bar j} \, \psi_R^{\dagger \bar j} \psi_L^{i})$
in the  classically vanishing anticommutator $\{Q_L,\, Q^\dagger_R\}$.
The above bifermion operator emerges as a quantum anomaly and 
acquires a nonperturbative vacuum expectation 
value of order of the scale parameter $\Lambda$ 
which determines the mass of the BPS kink.
Another well-known examples
of this type  are the $(1,0)$ central charges in ${\cal N}\!=1$ and ${\cal N}\!=2$
four-dimensional supersymmetric Yang--Mills (SYM) theories.
In the case of ${\cal N}\!=1$ the central charge  plays a crucial role
in domain walls \cite{DS}, and in ${\cal N}\!=2$ SYM  it gives the masses 
of all BPS states in the Seiberg--Witten solution.\footnote{\,This 
central charge anomaly was discussed in Ref. \cite{RNW4}}
 
 Anomaly in the central charges was extensively discussed in the case
 of two-dimensional Ginzburg--Landau models with minimal, ${\cal N}\!=1$,
 supersymmetry, see Ref. \cite{RNW2} and references therein.
These models are superrenormalizable. In contrast, the ${\cal N}\!=2$ 
CP$(N\!-\!1)$ models are logarithmic and in this respect much closer to
4D SYM.

As is well known, two-dimensional CP$(N\!-\!1)$ models 
allow an extension \cite{AGF} which preserves ${\cal N}\!=2$ supersymmetry
and introduces, in addition to $\Lambda$, 
free parameters $m_{a}$, the twisted masses. When the
twisted mass is much larger than 
$ \Lambda$ one can treat the model quasiclassically.
This provides a close parallel with the four-dimensional Seiberg--Witten analysis.
In fact, the two-dimensional CP(1) model with  twisted mass
was exactly solved \cite{Dor} in the same sense as the Seiberg--Witten 
solution of  ${\cal N}=2$ gauge theory in four dimensions. 
Among other consequences, examination of the exact solution
reveals the necessity of a bosonic anomalous term in the central charge.

In this paper we present a complete analysis of 
all anomalies, with emphasis on the central charge
anomaly, in two-dimensional sigma models with twisted masses.
First, we present the most general form of the ${\cal N}=2$ superalgebra
in two dimensions compatible with Lorentz invariance. 
Generally speaking, it could contain two complex central charges; only one of 
them appears in the CP$(N\!-\!1)$ model. Then we briefly summarize what was known 
previously of the central charges in this model. At $m=0$ the anomalous 
bifermion term was found in \cite{LS-03}. Analyzing Dorey's exact solution \cite{Dor},
valid at arbitrary $m$, in the quasiclassical limit of large $m$ 
we arrive at the conclusion that  an
additional bosonic operator in the central charge anomaly is inevitable.

Then we consider the conserved operators --- vector current, 
supercurrent, and the energy-momentum tensor --- 
which are combined in one supermultiplet. They enter different
components of the superfield ${\cal T}_\mu$, which we suggest to call
{\em hypercurrent} (instead of the term ``supercurrent'' used in
the literature). The current of the central charge also enters the hypercurrent.
We then derive the superconservation equation, the right-hand side of which contains 
the supermultiplet of all anomalies. Such an equation has been known in 
four-dimensional super-Yang--Mills theory since 1970s \cite{grisaru}.
Surprisingly, an analogous equation has never been derived
in two-dimensional ${\cal N}=2$ CP$(N\!-\!1)$ model.\footnote%
{\,For the minimal ${\cal N}=1$ supersymmetry in two dimensions 
the hypercurrent was treated in \cite{Shizuya}.}
 Here we close this gap.
The superfield equation explicitly demonstrates that
a single (one-loop) constant governs all anomalies.
Thus, it can be established from any of them.
In particular, we work out in detail a derivation whose  starting point is the
superconformal anomaly. It generalizes that of Ref.~\cite{LS-03}.
In principle, we could have stopped here. We carry out extra demonstrations, however.
Using various explicit ultraviolet regularizations (Pauli--Villars regularization,
higher derivatives) we calculate both the bosonic and bifermion terms 
in the central charge anomaly by virtue of a direct one-loop computation. 
Our result for the central charge successfully goes through a variety of checks:
renormalization-group analysis, compatibility with exact formulae \cite{Dor}, etc.

Finally, the last section of the paper treats an issue indirectly related to
the central charge problem. Namely, building on the results obtained in \cite{Dor}
we calculate the curve of marginal stability in CP(1).
This issue is of interest also due to its relation to the BPS sector in 4D ${\cal N}=2$ SQCD
with matter. We discuss  this relation.

\section{Sigma models with twisted mass}
\label{cpmwtm}
\setcounter{equation}{0}

Let us first briefly review the ${\cal N}=2$ supersymmetric sigma-models in 1+1 dimensions, $x^{\mu}=\{t,z\}$.
The target space is the $d$-dimensional K\"ahler manifold 
parametrized by the fields $\phi^{i},\,\phi^{\dagger\,\bar j}$, $\,i,\bar j=1,\ldots,d$,
which are the lowest components of the chiral and antichiral superfields 
$$
\Phi^{i}(x^{\mu}+i\bar \theta \gamma^{\mu} \theta),\qquad \Phi^{\dagger\bar j}(x^{\mu}-i\bar \theta \gamma^{\mu} \theta).
$$
With no twisted mass a generic Lagrangian of the ${\cal N}=2$ supersymmetric  sigma-model is \cite{WessBagger}
\begin{equation}
\label{eq:kinetic}
{\cal L}_{m=0}=\!\int\! {\rm d}^{4 }\theta K(\Phi, \Phi^{\dagger})
=G_{i\bar j} \big[\partial^\mu \phi^{\dagger\,\bar j}\, \partial_\mu\phi^{i}
+i\bar \psi^{\bar j} \gamma^{\mu} D_{\mu}\psi^{i}\big]
-\frac{1}{2}\,R_{i\bar jk\bar l}\,(\bar\psi^{\bar j}\psi^{i})(\bar\psi^{\bar l}\psi^{k})\,,
\end{equation}
where $K(\Phi, \Phi^{\dagger})$ is the K\"ahler potential,
$$
G_{i\bar j}=\frac{\partial^{2} K(\phi,\,\phi^{\dagger})}{\partial \phi^{i}\partial \phi^{\dagger\,\bar j}}
$$
 is the K\"ahler metric,
$R_{i\bar jk\bar l}$ is the Riemann tensor, 
$$ D_{\mu}\psi^{i}=
\partial_{\mu}\psi^{i}+\Gamma^{i}_{kl}\partial_{\mu} \phi^{k}\psi^{l}
$$
is the covariant derivative, and we use the notation
$\bar \theta=\theta^{\dagger}\gamma^{0}$, $\bar \psi=\psi^{\dagger}\gamma^{0}$
for the fermion objects.   The gamma-matrices  are chosen as 
\beq
\gamma^{0}=\gamma^t=\sigma_2\,,\qquad \gamma^{1}=\gamma^z = i\sigma_1\,,\qquad \gamma_{5} 
\equiv\gamma^0\gamma^1 = \sigma_3\,.
\label{sieeight}
\eeq

To deal with renormalizable models we limit our consideration to symmetric K\"ahler manifolds,
see Ref. \cite{Helgason} for definitions and classification.
For symmetric manifolds the Ricci-tensor $R_{i\bar j}$ is proportional to the metric,
\beq
\label{eq:RG}
R_{i\bar{j}} = \frac{g_{0}^2}{2}\,  b \, G_{i\bar{j}}\,.
\eeq
The coefficient  $b$ coincides with 
 the first (and the only) coefficient in the Gell-Mann--Low function.
The CP$(N\!-\!1)$ model is an example which we use as a reference point.\footnote{%
\,The  CP$(N\!-\!1)$ model is a special case, $n=N\!-\!1$, $m=1$,  of the Grassmann models with 
the symmetric K\"ahler manifold  ${\rm SU}(n+m)/{\rm SU}(n)\otimes {\rm SU}(m)\otimes {\rm U}(1)$.}
In this model
the target space is CP$(N\!-\!1)$ with $d=N\!-\!1$ coordinates and $b=N$.
For the massless CP($N\!-\!1)$ model 
a particular choice of the K\"ahler potential
\begin{equation}
\label{eq:kahler}
K_{m=0}=\frac{2}{g_{0}^{2}}\log\big(1+\sum_{i,\bar j=1}^{ N-1}\Phi^{\dagger\,\bar j}\delta_{\bar j i}\Phi^{i}\big)
\end{equation}
corresponds to the round Fubini--Study metric.

Let us  briefly remind how one can introduce the twisted mass parameters \cite{AGF, Dor}.
The theory \eqref{eq:kinetic} can be interpreted as an ${\cal N}=1$ theory of $d$ chiral superfields 
in four dimensions.  The theory possesses some number $r$ of U(1) isometries 
parametrized by $t^{a}$, $a=1,\ldots,r$.
The Killing vectors of the isometries can be expressed via derivatives of the Killing 
potentials $D^{a}(\phi, \phi^{\dagger})$,
\begin{equation}
\label{eq:KillD}
\frac{{d}\phi^{i}}{{  d}\,t_{a}}=-iG^{i\bar j}\,\frac{\partial D^{a}}{\partial \phi^{\dagger \,\bar j}}
\,,\qquad 
\frac{{d}\phi^{\dagger \,\bar j}}{{  d}\,t_{a}}=iG^{i\bar j}\,\frac{\partial D^{a}}{\partial \phi^{i}}\,.
\end{equation}
This defines U(1) Killing potentials up to additive constants.

In the case of  CP$(N\!-\!1)$ there are $N\!-\!1$ isometries 
 evident from the expression \eqref{eq:kahler} for the K\"ahler potential, 
\begin{equation}
\label{eq:iso}
\delta\phi^{i}=-i\delta t_{a} (T^{a})^{i}_{k}(\phi)^{k}\,,\qquad 
\delta\phi^{\dagger\,\bar j}=i\delta t_{a}(T^{a})^{\bar j}_{\bar l}\phi^{\dagger\,\bar l}\,,
\qquad a=1,\ldots, N-1\,,
\end{equation}
(together with the similar variation of fermionic fields),
where the  generators $T^{a}$ have a simple diagonal form,
\begin{equation}
(T^{a})^{i}_{k}=\delta^{i}_{a}\delta^{a}_{k}\,, \qquad a=1,\ldots,N-1\,.
\end{equation}
 The explicit form of the Killing potentials $D^{a}$ in CP$(N\!-\!1)$ with the Fubini--Study metric is
\beq
\label{eq:KillF}
D^{a}=\frac{2}{g_{0}^{2}}\,\frac{\phi^{\dagger}T^{a}\phi}{1+\phi^{\dagger}\phi}\,,
\qquad a=1,\ldots,N-1\,.
\eeq
Here we use the matrix notation implying that $\phi$ is a column $\phi^{i}$ and 
$\phi^{\dagger}$ is a row $\phi^{\dagger \bar j}$.

The isometries allow us  to introduce an interaction with $r$ {\em external} 
U(1) gauge 
superfields $V_{a}$ by modifying, in a gauge invariant way,  the K\"ahler potential \eqref{eq:kahler},
\begin{equation}
\label{eq:mkahler}
K_{m=0}(\Phi, \Phi^{\dagger})\to
K_{m}(\Phi, \Phi^{\dagger},V)\,.
\end{equation}
For CP$(N\!-\!1)$ this modification takes the form
\begin{equation}
\label{eq:mkahlerp}
K_{m}=\frac{2}{g_{0}^{2}}\log\big(1+\Phi^{\dagger}\,{\rm e}^{V_{a}T^{a}}\Phi\big)\,.
\end{equation}
In every gauge multiplet $V_{a}$ let us retain only the $A^{a}_{x}$ and $A^{a}_{y}$ 
components of the gauge potentials taking them to be just constants,
\beq
V_{a}=-m_{a}\bar \theta(1+\gamma_{5})\theta -\bar m_{a}\bar \theta(1-\gamma_{5})\theta\,,
\end{equation}
where we introduced complex masses  $m_{a}$ as linear combinations of 
constant U(1) gauge potentials,
\beq
m_{a}=A^{a}_{y}+iA^{a}_{x}\,,\qquad \bar m_{a}=m_{a}^{*}=A^{a}_{y}-iA^{a}_{x}\,.
\end{equation}
In spite  of the explicit $\theta$ dependence the introduction of masses does not 
break ${\cal N}=2$ supersymmetry.  One way to see this is to notice that the mass parameters 
can be viewed as the lowest components of the twisted chiral superfields
$D_{2}\bar D_{1}V_{a}$.

Now we can go back to two dimensions implying that there is no dependence
on $x$ and $y$ in the chiral fields.  It gives us the Lagrangian with the twisted masses 
included \cite{AGF, Dor}:
\begin{equation}
\label{eq:mtwist}
\begin{split}
{\cal L}_{m}&=\!\int\! {\rm d}^{4 }\theta \,K_{m}(\Phi, \Phi^{\dagger},V)
\\
&=G_{i\bar j}\, g_{MN}\big[{\cal D}^M\! \phi^{\dagger\,\bar j}\, {\cal D}^{N}\!\phi^{i}
+i\bar \psi^{\bar j} \gamma^{M}D^{N} \!\psi^{i}\big]
-\frac{1}{2}\,R_{i\bar jk\bar l}\,(\bar\psi^{\bar j}\psi^{i})(\bar\psi^{\bar l}\psi^{k})\,,
\end{split}
\end{equation}
where $G_{i\bar j} =\partial_{i}\partial_{\bar j}K_{m}|_{\theta=\bar\theta=0}$ is the K\"ahler metric 
and summation over $M$ includes, besides $M=\mu=0,1$, also 
$M=+,-$. 
The  metric $g_{MN}$ and extra gamma-matrices are
\begin{equation}
\label{eq:metric}
g_{MN}=\left(\begin{array}{crrr}1& 0& 0 & 0 \\0 & -1 & 0 & 0 \\[1mm]0 & 0 & 0 & -\frac 1 2 \\[1mm]0 & 0 & -\frac 1 2 & 0\end{array}\right),\qquad
\gamma^{+}=-i(1+\gamma_{5})\,,\quad
\gamma^{-}=i(1-\gamma_{5})\,.
\end{equation}
The gamma-matrices satisfy the following algebra:
\beq
\bar\Gamma^{M}\Gamma^{N}+\bar\Gamma^{N}\Gamma^{M}=2 g^{MN}\,,
\eeq
where the set $\bar\Gamma^{M}$ differs from $\Gamma^{M}$  by interchanging of
the $+,-$ components, $\bar\Gamma^{\pm}=\Gamma^{\mp}$.
The gauge covariant derivatives ${\cal D}^M$ are defined as
\begin{equation}
\begin{split}
{\cal D}^{\mu}\phi&=\partial^{\mu}\phi\,,\qquad {\cal D}^{+}\phi=-\bar m_{a}T^{a}\phi\,,
\qquad  {\cal D}^{-}\phi=m_{a}T^{a}\phi\,,
\\[1mm]
{\cal D}^{\mu}\phi^{\dagger}&=\partial^{\mu}\phi^{\dagger}\,,
\quad ~{\cal D}^{+}\phi^{\dagger}=\phi^{\dagger} T^{a}\bar m_{a}\,,
\qquad  {\cal D}^{-}\phi^{\dagger}=-\phi^{\dagger} T^{a} m_{a}\,,
\end{split}
\end{equation}
and similarly for ${\cal D}^{M}\psi$, while the general covariant derivatives $D^{M}\psi$ are
\begin{equation}
\begin{split}
D^{M}\psi^{i}=
{\cal D}^{M}\psi^{i}+\Gamma^{i}_{kl}\,{\cal D}^{M}\! \phi^{k}\,\psi^{l}\,.
\end{split}
\end{equation}

Let us present explicit expressions in the case  of CP(1). 
In this case a single complex field $\phi(t,z)$ serves as 
coordinate on the target space which is equivalent to $S^{2}$.
The K\"ahler and Killing potentials, $K$ and $D$, the metric $G$, the Christoffel symbols $\Gamma,\,\bar\Gamma$ and the Ricci tensor $R$ are then 
\beq
\begin{split}
~~~~&K_{m}\big|_{\theta=\bar\theta=0}=\frac{2}{g_{0}^{2}}\,\log \chi\,,
\quad 
D=\frac{2}{g_{0}^{2}}\,\frac{\phi^{\dagger}\phi}{\chi}\,,
\quad
G=G_{1\bar 1}=\partial_\phi\partial_{\phi^\dagger\,} K_{m}\big|_{\theta=\bar\theta=0}=
\frac{2}{g_{0}^2\,\chi^{2}}\,,\qquad\qquad
\\[2mm]
&\Gamma =\Gamma^{1}_{11} =- 2\, \frac{\phi^\dagger\,}{\chi}\,,\quad 
\bar\Gamma =\Gamma^{\bar 1}_{\bar 1\bar 1}=
- 2\, \frac{ \phi}{\chi}\,,\quad
 R = R_{1\bar 1}=-G^{-1}\!R_{1\bar 1 1\bar 1}=\frac{2}{\chi^2}\,,
 \end{split}
\label{Atwo}
\eeq
where we use the notation 
\beq
\chi \equiv 1+\phi\,\phi^\dagger\,.
\eeq
The Lagrangian of the 
CP(1) model takes the 
following form \cite{AGF}:
\begin{equation}
\begin{split}
{\cal L}_{\,  CP(1)}=
G&\,\Big\{{\cal D}_M \phi^{\dagger}\, {\cal D}^M \phi+i\bar \psi \gamma^{M}D_{M}\psi
+\frac{R}{2}(\bar \psi \psi)^{2}\Big\}
\\[1mm]
= G&\,\Big\{
\partial_\mu \phi^{\dagger}\, \partial^\mu\phi -|m|^{2} \, { \phi^{\dagger}\,\phi}
+i\bar \psi \gamma^{\mu}\partial_{\mu}\psi -\frac{1- \phi^{\dagger}\,\phi}{\chi} \,\bar \psi\,\mu\,
\psi
\\[1mm]
&
-\frac{2i}{\chi}\, \phi^{\dagger}\partial_{\mu}\phi\,\bar\psi \gamma^{\mu}\psi+
\frac{1}{\chi^{2}}\,(\bar \psi \psi)^{2}\Big\}\,,
\end{split}
\label{cpone}
\end{equation}
where 
\beq
\mu=m\,\frac{1+\gamma_{5}}{2}+\bar m \,\frac{1-\gamma_{5}}{2}\,.
\label{eq:mmu}
\eeq
One can also add the  $\theta$  term
$$
\frac{i g_{0}^{2}\,\theta}{4\pi}\,G\,
\varepsilon^{\mu\nu} \partial_\mu \phi^{\dagger}\,\partial_\nu \phi \,,
$$
which is a total derivative, to the Lagrangian  \eqref{cpone}.
The vacuum angle $\theta$ enters  physics in the combination
$\theta+2 \arg m$, where $\arg m$ is the phase of the complex mass $m$,
so we can safely include $\theta$ into this phase.
 
In terms of components of $\psi$,
\beq
\psi =\left(\begin{array}{cc}
\psi_R \\
\psi_L
\end{array}
\right),
\eeq
the Lagrangian \eqref{cpone} can be rewritten as
\beq
\begin{split}
{\cal L}_{\,  CP(1)}&= G\, \Big\{
\partial_\mu \phi^\dagger\, \partial^\mu\phi -|m|^2{\phi^\dagger\,\phi}
+\frac{i}{2}\big(\psi_L^\dagger\!\stackrel{\leftrightarrow}{\partial_R}\!\psi_L 
+ \psi_R^\dagger\!\stackrel{\leftrightarrow}{\partial_L}\!\psi_R
\big)
\\[1mm] 
&
-i\,\frac{1-\phi^\dagger\,\phi}{\chi} \,\big(m\,\psi_L^\dagger \psi_R + \bar m
\psi_R^\dagger \psi_L
\big)
-\frac{i}{\chi}\,  \big[\psi_L^\dagger \psi_L
\big(\phi^\dagger \!\stackrel{\leftrightarrow}{\partial_R}\!\phi
\big)+ \psi_R^\dagger\, \psi_R
\big(\phi^\dagger\!\stackrel{\leftrightarrow}{\partial_L}\!\phi
\big)
\big]
\\[1mm]
&
-
\frac{2}{\chi^2}\,\psi_L^\dagger\,\psi_L \,\psi_R^\dagger\,\psi_R
\Big\}\,,
\end{split}
\label{Aone}
\eeq
where 
\beq
\partial_L =\frac{\partial}{\partial t} +\frac{\partial}{\partial z}\,,\qquad
\partial_R =\frac{\partial}{\partial t} - \frac{\partial}{\partial z}\,.
\eeq

\section{Superalgebra}
\label{supa}
\setcounter{equation}{0}

The most general form of the centrally extended algebra for four supercharges $Q_{\alpha}$, 
$Q^{\dagger}_\beta$ consistent with Lorentz symmetry in 1+1 dimensions is
\beq
\begin{split}
&\big\{Q_{\alpha}, Q_{\beta}^{\dagger}\big\}(\gamma^{0})_{\beta\gamma}=2\left[ P_{\mu}\gamma^{\mu}
+ {Z}\,\frac{1-\gamma_{5}}{2} +{Z}^\dagger\,\frac{1+\gamma_{5}}{2}\right]_{\alpha\gamma},
\\[1mm]
&\big\{Q_{\alpha},Q_{\beta}\big\}(\gamma^{0})_{\beta\gamma}=-
2 {Z'}\,(\gamma_{5})_{\alpha\gamma}\,,\qquad
\big\{Q_{\alpha}^{\dagger}, Q_{\beta}^{\dagger}\big\}(\gamma^{0})_{\beta\gamma}=
2 {Z'}^{\dagger}\,(\gamma_{5})_{\alpha\gamma}\,.
\end{split}
\label{eq:supalg}
\eeq
The algebra contains two complex central charges, $Z$ and $Z'$. 
In terms of  components $Q_{\alpha}=(Q_{R}, Q_{L})$ the nonvanishing anticommutators are
\beq
\begin{split}
&\{ Q_L , Q_L^\dagger\} = 2(H+ P)\,,\qquad \{Q_R, Q_R^\dagger    \} = 2(H-P)\,,
\\[1mm]
&\{Q_L, Q_R^\dagger  \} =2i{Z}\,,\qquad\qquad
~~\{Q_R, Q_L^\dagger \}= -2i {Z}^{\dagger}\,,
\\[1mm]
&\{Q_L, Q_R \} =2i{Z'}\,,\qquad\qquad
~\{Q_R^{\dagger}, Q_L^\dagger \}= -2i {Z'}^{\dagger}\,.
\end{split}
\label{compalg}
\eeq
It exhibits the automorphism
$Q_{R}\leftrightarrow Q^{\dagger}_{R},\, Z\leftrightarrow Z'$  associated \cite{Dor} with the transition to a mirror representation \cite{HH}. 

The superalgebra  \eqref{eq:supalg} leads to the constraint
\beq
M\geq |Z|+|Z'|
\eeq
for the particle masses. The bound is saturated by 1/4 BPS states when both 
$Z$ and $Z'$ are nonvanishing; when one of the central charges vanishes we deal with 
1/2 BPS states.

In the models \eqref{eq:mtwist} with the twisted mass 
one can use canonical quantization to determine  the central charges.
In the classical approximation, i.e. without
anomalous contribution from quantum loops,
the central charge  $Z'$ vanishes and  $Z$ takes the form \cite{Dor,LS-03},
\beq
Z=
\sum_{a=1}^{r} m_{a} \,q^{a}- i\int\! dz\, \partial_z\, O \,.
\label{seventeen}
\eeq
Here  $q^{a}$
are the charges of the global U(1) symmetries of the model, Eq.\,\eqref{eq:iso},
\beq
q^{a} \equiv   \int dz \, {\cal J}_0^{a}\,,
\eeq
where the Noether currents in the CP$(N\!-\!1)$ case are
\beq
{\cal J}_\mu^{a} = G_{i\bar j}\Big[ \phi^{\dagger\bar j}T^{a}
i\! \stackrel{\leftrightarrow}{\partial_{\mu}}\!\phi^{i} +
\bar\psi^{\bar j} T^{a}\gamma_\mu\big(\psi^{i}+\Gamma^{i}_{lk}\phi^{l}\psi^{k}\big)
\Big]\,.
\label{fourteen}
\eeq
The second term in Eq.\eqref{seventeen} clearly represents a topological charge.
The operator $O$ is a local operator; its classical part is given by 
the Killing potentials $D^{a}$,
\beq
\label{eq:canO}
O_{\rm canon}=\sum_{a=1}^{r}m_{a}D^{a}(\phi, \phi^{\dagger})\,.
\eeq
Let us also  introduce the current of the central charge $\zeta_{\mu}$,
\beq
\label{eq:centcur}
\zeta_{\mu}=m_{a}{\cal J}_\mu^{a}-i\varepsilon_{\mu\nu}\partial^{\nu}O\,,
\qquad Z=\int\! dz \,\zeta_{0} \,.
\eeq

To determine the loop corrections to the topological central charges which are integrals of the total derivatives
it is convenient to consider instead of the anticommutators of the supercharges 
their anticommutators with the local supercurrent.  
In the tree-level approximation the supercharges  are presented as
\beq
\begin{split}
&Q_{\alpha}=\int dz J^{0}_{\alpha}\,,
\\[1mm]
&J^{\mu}_{\alpha}=\sqrt{2}\,G_{i\bar j} \big[{\cal D}_{N}\phi^{\dagger\bar j} \,\bar\Gamma^{N}\gamma^{\mu}\psi^{i}\big]_{\alpha}
=\sqrt{2}\,G_{i\bar j}\big[\partial_{\nu}\phi^{\dagger\bar j} \gamma^{\nu}\gamma^{\mu}\psi^{i}
+i\phi^{\dagger\bar j}\gamma^{\mu}\mu_{a}\,\psi^{i}\big]_{\alpha}\,.
\end{split}
\label{eq:superJ}
\eeq
where we use the notation 
$\mu_{a}=m_{a}(1+\gamma_{5})/2+\bar m_a (1-\gamma_{5})/2$.
Consider the anticommutator of the supercurrent $J^{\mu}_{\alpha}$
and supercharge $\bar Q_{\beta}$. The canonical commutation relations lead to
\beq
\label{eq:JQ}
\left\{J_{\mu,\alpha}\,,\bar Q _\beta\right\}_{\rm canon} =2\Big\{
\gamma^\nu\vartheta_{\mu\nu}
-\frac{i}{2}\,{\not\!  \partial}\, {\cal V}_{\mu}
+\frac{1-\gamma_{5}}{2}\,\zeta_{\mu}+\frac{1+\gamma_{5}}{2}\,\zeta_{\mu}^{\dagger}
 \Big\}_{\alpha\beta}\,,
\eeq
where $\vartheta_{\mu\nu}$ is the energy-momentum tensor and
${\cal V}_\mu$  is the vector fermionic current,
\beq
{\cal V}^\mu=G_{i\bar j}\bar\psi^{\bar j}\gamma^{\mu}\psi^{i}\,.
\eeq
Note that at  tree level this current is algebraically related to
the axial current ${\cal A}_{\mu}$,
\beq
{\cal A}_{\mu}= -\varepsilon_{\mu\nu}{\cal V}^{\,\nu}
=G_{i\bar j}\bar\psi^{\bar j}\gamma_{\mu}\gamma_{5}\psi^{i}\,. 
\eeq
The current of the central charge $\zeta_{\mu}$ is defined in Eq.\,\eqref{eq:centcur}.
Its topological part is expressed via a local operator $O$ whose classical part
is given in Eq.\,\eqref{eq:canO}. 

In Secs.\ 5-8 we will calculate  the quantum part of the operator $O$
which represent the anomalous contribution to the central charge $Z$,
 \beq
 \label{eq:Oanom}
 O_{\rm anom}=
 -\frac{g_{0}^{2}b}{4\pi}\Big(\sum_{a}^{r} m_{a} D^{a}+ G_{i\bar j}\,\bar \psi^{\bar j}\,
\frac{1-\gamma_{5}}{2}\,\psi^{i}\Big)\,,
 \eeq
where $b$ stands for the first
 coefficient in the Gell-Mann--Low function. In the \mbox{CP$(N\!-\!1)$} case
$b=N$ and $r=N\!-\!1$. 
The relation \eqref{eq:RG} between the Ricci tensor and the metric 
 allows to rewrite Eq.\,\eqref{eq:Oanom} as
\beq
 \label{eq:Oanom1}
 O_{\rm anom}=
 -\frac{g_{0}^{2}b}{4\pi}\sum_{a}^{r} m_{a} D^{a}- \frac{1}{4\pi}R_{i\bar j}\,\bar \psi^{\bar j}
(1-\gamma_{5})\psi^{i}\,.
 \eeq
No anomaly
appears in $Z'$, so this central charge vanishes both at the classical and quantum levels.

The anomalous part of the topological current enters the supermultiplet of anomalies;
other entries are the divergence of the axial current  $\partial_{\mu}{\cal A}^{\mu}$,
the superconformal anomaly in the supercurrent
$\gamma^{\mu}J_{\mu}$, and conformal (or dilatational) anomaly
in the trace of the energy-momentum tensor $\vartheta^{\mu}_{\mu}$. 
All these anomalies will be determined too.

\section{How do we learn of the existence of the central charge anomaly?}
\label{hdwlotcca}
\setcounter{equation}{0}

In the next section we will construct the supermultiplet of anomalies using 
the superfield description.
In fact, the anomalous bifermion term in  the central charge, see Eq.\,\eqref{eq:Oanom},
was derived in \cite{LS-03} in the $m=0$ case from the superconformal anomaly
using the ${\cal N}=2$ supersymmetry of the model.
That the bifermion operator must be accompanied at $m\neq 0$ by
a pure bosonic one follows from supersymmetry together with gauge invariance,
as we will see 
in the next section. It also follows from 
renormalization properties at one loop (see Section \ref{sec:super}).

The occurrence of the anomaly in the central charge can be seen from the following argument.
In the CP(1) model with twisted mass
an exact expression for the central charge in the case of the soliton 
carrying a nontrivial topologic quantum number
 was obtained by Dorey \cite{Dor} on the basis of a mirror formula of
Hanany--Hori \cite{HH} (see also the discussion at end of this Section), 
\beq
\big\langle Z \big\rangle_{{\rm kink}}= -\frac{i}{2\pi}
\left\{m\log\frac{m+\sqrt{m^2+4\Lambda^2} }
{m- \sqrt{m^2+4\Lambda^2}} -
2\sqrt{m^2+4\Lambda^2}\right\}.
\label{ndef}
\eeq
Note the infinite multivaluedness associated with
branches of the logarithm, in addition to the square-root
branches reflecting the vacuum structure (two vacua in the CP(1) model).
Resulting structure of monodromies provides an extra check 
of the expression \eqref{seventeen} for the central charge:
every $2\pi$ rotation in the complex plane of $m^{2}$ at large $m^{2}$ shifts 
the U(1) charge $q$ in \eqref{seventeen} by one unit,
and also changes the sign of
the  topological charge, transforming soliton into antisoliton.
 
\vspace{2mm}

Let us consider the quasiclassical limit of Eq.~(\ref{ndef}) when the mass $m$ is real and large, $ m\gg\Lambda$.
In this limit
\beqn
\big\langle Z \big\rangle_{{\rm kink}} \! &=&\!-\frac{i\, m}{2\pi}\left[\log\Big(\!-\frac{m^2}{\Lambda^2}\Big)-2
\right]\nonumber\\[2mm]
&=&\! \frac{1}{2}\,m -im\left(\frac{2}{g_0^2} -\frac{1}{2\pi}\log \frac{M_0^2}{m^2}\right) +\frac{i\,m}{\pi}\,,
\label{241}
\eeqn
where $g^2_0$ is the bare coupling constant, and $M_0$ is the
ultraviolet cut off. The first term in the second line reflects the fractional U(1) charge, 
$q =1/2$, carried by the soliton.\footnote{%
\,The reason for the occurrence of half-integer charge is explained in detail in the lecture 
\cite{Shifman}.}
 The second term coincides with the one-loop
corrected average of $(\!-i\!\int\!\! dz \partial_{z}O_{\rm canon})$ in the central charge.
The third term $im/\pi$ represents the anomaly.

Indeed, one can readily check that
the charge renormalization for $4\pi/g_{0}^{2}$ is given by $\log M_0^2/m^2$, with no nonlogarithmic constant.
The same statement applies to the one-loop correction in $O_{\rm canon}$,
which identically coincides with the charge renormalization.
The presence of the nonlogarithmic constant in Eq.~(\ref{241})
demonstrates the need for the bosonic term $-mg_{0}^{2}D/2\pi$ in the central charge anomaly
 \beq
O_{\rm anom} = -\frac{1}{2\pi}\big(m g_{0}^{2} D -iR\, \psi_R^\dagger\,\psi_L
\big)\,.
\label{245}
\eeq
The necessity to have also the bifermion part of the anomaly can be seen in 
the semiclassical approach from the absence of higher loops in 
the expression (\ref{241}); it contains only the one-loop term. 
Indeed, without the bifermion term the operator $O_{\rm anom}$ 
is not renormalization-group invariant, the first loop brings in the factor $1-(g_{0}^{2}/4\pi)\log M_{0}^{2}/m^{2}$
in $\langle D\rangle$, as we discussed above. The logarithmic mixing of $\psi_R^\dagger\,\psi_L$ with $D$ 
leads to cancellation of the logarithms and makes   $O_{\rm anom}$  renormalization-group invariant.
 
 To conclude this section we would like to note, for completeness, that:

\noindent
(i) The matrix element of the
operator $O_{\rm anom}$ in the central charge anomaly (more exactly, the difference
of its vacuum averages in two vacua of the model at hand) can be found exactly,
\beq
  \Delta \,\big\langle O_{\rm anom} 
  \big\rangle =-\frac{1}{\pi}\,\sqrt{m^2+4\Lambda^2} \,.
\label{246}
\eeq

\noindent
(ii)
The above-mentioned mirror representation \cite{HH} from which Eq.~(\ref{ndef})
ensues is quite  straightforward in the case under consideration.  For the 
CP(1) model with twisted mass  the mirror 
representation introduces a superpotential,
\beq
{\cal W} = \frac{i}{2\pi}\,\Big\{ z +\frac{\Lambda^2}{z} - m\log z
\Big\}\,,
\label{certwo}
\eeq
where $z$ is a chiral superfield. This superpotential has
two critical points, where $\partial{\cal W}/\partial z |_{z=z_{\pm}}=0$,
$$ z_{\pm} = \frac{m\pm \sqrt{m^2+4\Lambda^2}}{2}\,.$$
The central charge (\ref{ndef}) is given then by ${\cal W}(z_{+})-{\cal W}(z_{-})$.

We will return to  the BPS solitons in Section \ref{cms} to discuss their spectrum
as a function of the complex parameter $m$, particularly, the  issue of 
the curve of marginal stability.

\section{Supermultiplets of currents and anomalies}
\label{sec:supermult}
\setcounter{equation}{0}

The anticommutator (\ref{eq:JQ}) demonstrates that  the supercurrent $J_{\mu}$ and 
energy-mo\-men\-tum tensor $\vartheta_{\mu\nu}$ enter the same supermultiplet, together
with the fermion current ${\cal V}_{\mu}$ and the current of the central charge $\zeta_{\mu}$.
All these objects can be viewed as  
different components of one and the same superfield ${\cal T}_{\mu}$,
let us call it hypercurrent,\footnote{%
\,Often this superfield  is called supercurrent but  we use this term for $J_{\mu}$.}
\begin{equation}
{\cal T}_{\mu}={\cal V}_{\mu}+\big[ \theta \gamma^{0}J_{\mu} +{\rm h.c.}\big]
-2\bar\theta \gamma^{\nu}\theta\, \vartheta_{\mu\nu}+ 
\big[\bar\theta(1+ \gamma_{5})\theta \zeta_{\mu}+{\rm h.c.}\big]+\ldots \,.
\end{equation}
In the subsequent equations we refer to the CP(1) case -- the generalization is straightforward.

In CP(1) the hypercurrent ${\cal T}_{\mu}$ can be written as
\beq
{\cal T}^{\mu}=\frac{1}{2}\,(\gamma^{0}\gamma^{\mu})_{\beta\alpha}{\cal T}^{\beta\alpha}\,,
\qquad {\cal T}^{\beta\alpha}=G\bar D^{\beta}(\Phi^{\dagger}{\rm e}^{V})\,{\rm e}^{-V}D^{\alpha}({\rm e}^{V}\Phi)\,,
\eeq
where $D_{\alpha}$, $\bar D_{\beta}$ are conventional spinor derivatives,
and the metric $G= \partial^{2} K_{m}/\partial \Phi  \,\partial\Phi^{\dagger}$ 
should be viewed as a superfield.
Only two  components of ${\cal T}^{\beta\alpha}$ are relevant to the 
hypercurrent ${\cal T}^{\mu}$. With our choice \eqref{sieeight} of $\gamma$ 
matrices they are  ${\cal T}^{11}={\cal T}^{0}+{\cal T}^{1}$ and
${\cal T}^{22}={\cal T}^{0}-{\cal T}^{1}$. 

At the classical level 
\begin{equation}
\label{eq:class}
\bar D_{1}{\cal T}^{11}\big|_{\rm class}=-2i\bar m\bar D^{1}\tilde D\,,\qquad
\bar D_{2}{\cal T}^{22}\big|_{\rm class}=-2i m\bar D^{2}\tilde D\,,
\end{equation}
where
\begin{equation}
\tilde D=\frac{2}{g_{0}^{2}}\,\frac{\Phi^{\dagger}{\rm e}^{V}\Phi}{1+\Phi^{\dagger}{\rm e}^{V}\Phi}
\end{equation}
is the superfield generalization of the Killing potential, see Eq.\,\eqref{eq:KillF}.
Applying the Hermitean conjugation we get similar equations for $D_{1}{\cal T}^{11}$ and $D_{2}{\cal T}^{22}$.

At the quantum level the one-loop anomalies in the twisted CP(1) modify the 
right-hand side of Eq.\,\eqref{eq:class},
\begin{equation}
\label{supanom1}
\begin{split}
\bar D_{1}{\cal T}^{11}=-2i\bar D^{1}\Big\{\bar m\tilde D- \frac{g_{0}^{2}}{4\pi}\,iD^{\alpha}\bar D_{\alpha}K\Big\}
\,,\\[2mm]
\bar D_{2}{\cal T}^{22}=-2i\bar D^{2}\Big\{m\tilde D- \frac{g_{0}^{2}}{4\pi}\,iD^{\alpha}\bar D_{\alpha}K\Big\}
\,.
\end{split}
\end{equation}
The coefficient of  $D^{\alpha}\bar D_{\alpha} K$ in the anomalous part is
fixed by the trace anomaly $\vartheta^{\mu}_{\mu}$\,,
\beq
\label{eq:trace1}
\big(\vartheta_{\mu}^{\mu}\big)_{\rm anom}=
 \frac{g_{0}^{2}}{2\pi}\,G\big(\partial_{\mu}\phi^{\dagger}\partial^{\mu}\phi-|m|^{2}\phi^{\dagger}\phi\big)
 \,,
 \eeq
which follows from the one-loop $\beta$ function.  In Eq.\,\eqref{supanom1} this anomaly enters  the component linear in $\theta$.

The generalization of  \eqref{supanom1} to an arbitrary symmetric K\"ahler target space is straightforward:
\begin{equation}
\label{supanom}
\begin{split}
\bar D_{1}{\cal T}^{11}=-2i\bar D^{1}\Big\{\sum_{a}\bar m_{a}\tilde D^{a}- \frac{bg_{0}^{2}}{8\pi}\,iD^{\alpha}\bar D_{\alpha}K\Big\}
\,,\\[2mm]
\bar D_{2}{\cal T}^{22}=-2i\bar D^{2}\Big\{\sum_{a} m_{a}\tilde D^{a}- \frac{bg_{0}^{2}}{8\pi}\,iD^{\alpha}\bar D_{\alpha}K\Big\}
\,.
\end{split}
\end{equation}
Note that  the superfield $D^{\alpha}\bar D_{\alpha} K=(D_{2}\bar D_{1}-D_{1}\bar D_{2})K$ which contains all anomalies
is the difference of the twisted chiral superfield $-D_{1}\bar D_{2}K$ and its complex conjugate,
the twisted antichiral field $-D_{2}\bar D_{1}K$\,. Only the antichiral (chiral) part contributes in the first (second) line of Eq.\,\eqref{supanom}.

Consistency of the above equations with Lorentz symmetry is clear because 
the anomalous addition  $D^{\alpha}\bar D_{\alpha} K$  to the Killing 
potential  is a Lorentz scalar. To demonstrate consistency with reparametrization  invariance in the target space we can rewrite Eq.\,\eqref{supanom} as
\begin{equation}
\label{supanom1p}
\begin{split}
\bar D_{1}{\cal T}^{11}=-2i\bar D_{2}\Big\{\Big(1-\frac{bg_{0}^{2}}{4\pi}\Big)\sum_{a}\bar m_{a}\tilde D^{a}+ \frac{ibg_{0}^{2}}{8\pi}\,{\cal T}^{21}\Big\}
\,,\\[2mm]
\bar D_{2}{\cal T}^{22}=2i\bar D_{1}\Big\{\Big(1-\frac{bg_{0}^{2}}{4\pi}\Big) \sum_{a}m_{a}\tilde D^{a}- \frac{ibg_{0}^{2}}{8\pi}\,{\cal T}^{12}\Big\}
\,.
\end{split}
\end{equation}
The above equations contain all anomalies.  In particular, the lowest component 
of the braces in the second equation in \eqref{supanom1} gives $O_{\rm canon}+O_{\rm anom}$
for the central charge $Z$, see Eqs.\,\eqref{seventeen}, \eqref{eq:canO} and \eqref{eq:Oanom}.
More details about the component form of different anomalies are given in the next section.

It is interesting to compare the results above with the superfield description
of anomalies in 4D super Yang--Mills (SYM) theory (for a review see \cite{grisaru}). 
In ${\cal N}\!=\!1$ SYM the hypercurrent 
\begin{equation}
{\cal T}_{\mu} = -\frac{2}{g^2}\,(\sigma_\mu)^{\alpha\dot\alpha}\mbox{Tr} \left[{\rm e}^V
W_\alpha {\rm e}^{-V}\bar W_{\dot \alpha}\right]
= R_{\mu} \!-
\left[  i\theta^{\alpha} J_{\mu\alpha} 
\! + \!
\mbox{h.c.}    \right] -
 2\, \theta\sigma^{\nu}\bar{\theta}  \,
     \vartheta_{\mu\nu}\! +\!\dots\;,
\label{decom2}
\end{equation}
contains the axial current $R_{\mu}=(1/g^{2})\lambda \sigma_{\mu}\bar \lambda$ (as its lowest
component) together with the supercurrent and energy-momentum tensor.  All anomalies 
are collected in the relation
\begin{equation}
\bar{D}_{\dot\alpha}{\cal T}^{\alpha\dot\alpha} =
\frac{b}{24\pi^2}\,D^\alpha\, {\rm Tr}\,W^2\,,
\label{sganom}
\end{equation}
where $b$ is the first coefficient in the Gell-Mann--Low function in SYM.
The similarity with Eq.\,\eqref{supanom1} is clear.
There is no classical part in the case of  SYM. The chirality of spinor derivatives 
is different on the opposite sides of Eq.\,\eqref{sganom} but not in Eq.\,\eqref{supanom1}.
This distinction goes away if one passes
to the twisted superfield chirality in Eq.\,\eqref{supanom1}.
A real distinction refers
to $\partial_{\mu} {\cal T}^{\mu}$ which vanishes for sigma models
but not for SYM  \cite{grisaru}.

\section{The central charge anomaly from the superconformal anomaly}
\label{sec:super}
\setcounter{equation}{0}

In the case of CP(1), the full expression for the  central charge in the algebra 
(\ref{seventeen}) is\,\footnote{%
\,The factor $i$ in front of $R\,\psi_R^\dagger\,\psi_L$ is missing in Eq.\,(10.9) of 
Ref.~\cite{LS-03}.}
\beq
Z= mq -i\!\int\! dz \,\partial_{z}
\Big\{m \,D - \frac{1}{2\pi}\big( m g_{0}^{2} D
-i\, R\,\psi_R^\dagger\,\psi_L\big)
\Big\},
\label{cerfive}
\eeq
where $q$ stands for the Noether charge of the global U(1).
In this section we will derive this expression using supersymmetry
of the model to connect it to the superconformal anomaly $\gamma^\mu J_\mu$.
This derivation extends that of Ref.~\cite{LS-03}. Simultaneously we will 
get all other anomalies.

The covariant expression for the supercurrent $J_{\mu}$ is given in Eq.\,\eqref{eq:superJ}.
Contracting this conserved supercurrent  with  $\gamma^{\mu}$ we get in $D=2-\varepsilon$ dimensions
\beq 
\begin{split}
&\gamma^{\mu}J_{\mu}=2\sqrt{2} i G \mu\phi^{\dagger}\psi+ 
(\gamma^{\mu}J_{\mu})_{\rm anom}\,,
\\[2mm]
&(\gamma^{\mu}J_{\mu})_{\rm anom} = \sqrt{2} \,\varepsilon\, G\left[ 
\gamma^{\nu}\partial_{\nu}\phi^{\dagger}\psi -i\mu\phi^{\dagger}\psi \right].
\end{split}
\eeq
The vanishing of $\gamma^{\mu}J_{\mu}$ as $\varepsilon \to 0$ and $\mu\to 0$ corresponds to the 
classical superconformal invariance. It is well known that this symmetry is anomalous. The conformal anomaly manifests itself  through the coupling constant renormalization
\begin{equation}
\label{eq:cren}
\frac{1}{g_0^2} = \frac{1}{g^2} + \frac{b}{4\pi}\, \frac{1}{\varepsilon} \, ,
\end{equation}
which cancels the factor $\varepsilon$. As we discussed earlier  the first
 coefficient in the Gell-Mann--Low function 
$b=N$ for CP$(N\!-\!1)$ model. Moreover, Eq.\,\eqref{eq:RG}
which relates the Ricci tensor and the metric 
allows us to rewrite the anomalous part of $\gamma^{\mu}J_{\mu}$ as
\beq
(\gamma^{\mu}J_{\mu})_{\rm anom} = \frac{R}{\sqrt{2}\pi} \left[ 
\gamma^{\nu}\partial_{\nu}\phi^{\dagger}\psi -i\mu\phi^{\dagger}\psi \right]
= \frac{R}{\sqrt{2}\pi} \,\Gamma^{N}{\cal D}_{N}\phi^{\dagger}\psi
\,.
\eeq

Let us now calculate  the anticommutators of $(\gamma^{\mu}J_{\mu})_{\rm anom}$
with the supercharges $Q$ and $\bar Q$.
To this end we use the action of the supercharges on the fundamental fields
collected in the Appendix, see Eqs.\,(\ref{eq:a3}), (\ref{eq:a5}). 
For the anticommutator with $Q$ we get 
\beq
\left\{(\gamma^{\mu}J_{\mu})^{\rm anom}_{\alpha}, Q _\beta\right\} =0\,.
\eeq
This shows that there is no one-loop quantum correction in the anticommutator 
$\{Q_{\alpha},Q_{\beta}\}$. Thus, the central charge $Z'$ remains vanishing.

Commuting $(\gamma^{\mu}J_{\mu})_{\rm anom}$ with $\bar Q$ we arrive at
\beq
\begin{split}
\left\{(\gamma^{\mu}J_{\mu})^{\rm anom}_{\alpha},\bar Q _\beta\right\} &=2\Big\{\,
\frac{R}{2\pi}\big(\partial_{\mu}\phi\partial^{\mu}\phi-|m|^{2}\phi^{\dagger}\phi\big)
\\[2mm]
&+\gamma_{5}\,\frac{1}{4\pi}\,\big[2R\varepsilon^{\mu\nu}\partial_{\mu}\phi^{\dagger}\partial_{\nu}\phi
+i\partial_{\mu}(R\bar\psi\gamma^{\mu}\gamma_{5}\psi)\big]
\\[2mm]
&-i\,\frac{R}{2\pi}\,\mu\gamma^{\mu}\partial_{\mu}(\phi^{\dagger}\phi)
-\frac{i}{4\pi}\,\gamma^{\mu}\big[\partial_{\mu}(R\bar\psi \psi)+\varepsilon_{\mu\nu}\partial^{\nu}
(R\bar\psi \gamma_{5}\psi)\big]\Big\}_{\alpha\beta}
\,.
\end{split}
\label{eq:anomQ}
\eeq
Compare  this result   with the general expression  \eqref{eq:JQ},
\beq
\label{eq:corr}
\begin{split}
\left\{J_{\mu,\alpha}\,,\bar Q _\beta\right\} =2\Big\{
\gamma^\nu\vartheta_{\mu\nu}
-\frac{i}{2}\,{\not\!  \partial} \,{\cal V}_{\mu}
+ \mu^{\dagger}\big[{\cal J}_\mu+i\gamma_{5}\varepsilon_{\mu\nu}\partial^{\nu}D\big]
\\[2mm]
-i\varepsilon_{\mu\nu}\partial^{\nu}\Big[\frac{1-\gamma_{5}}{2}\, O_{\rm anom}
-\frac{1+\gamma_{5}}{2}\, O^{\dagger}_{\rm anom}\Big]
 \Big\}_{\alpha\beta}
 \,,
 \end{split}
\eeq
where 
terms in the second line account for a loop modification of the central charge 
current $\zeta_{\mu}$.
 No other modifications are allowed because of conservation of 
$J_{\mu}$, $\vartheta_{\mu\nu}$, ${\cal V}_{\nu}$ and ${\cal J}_\mu$\,; only the topological part could be modified.
Convoluting \eqref {eq:corr} with $(\gamma^{\mu})_{\gamma\alpha}$ and retaining only the anomalous part we arrive at
\beq
\label{eq:anomQ1}
\begin{split}
\left\{(\gamma^{\mu}J_{\mu})^{\rm anom}_{\gamma},\bar Q _\beta\right\} 
=2\Big\{
\big(\vartheta_{\mu}^{\mu}\big)_{\rm anom}
-\frac{i}{2}\,\gamma_{5}\big(\partial^{\mu}{\cal A}_{\mu}\big)_{\rm anom}
\\[2mm]
 -i\gamma^{\mu}\varepsilon_{\mu\nu}\partial^{\nu} 
 \Big[\frac{1-\gamma_{5}}{2}\, O_{\rm anom}
-\frac{1+\gamma_{5}}{2}\, O^{\dagger}_{\rm anom}\Big]
 \Big\}_{\gamma\beta}\,,
 \end{split}
\eeq
where we use  the axial current ${\cal A}_{\mu}=-\varepsilon_{\mu\nu}{\cal V}^{\nu}$. 

Comparing terms with the unit matrix in Eqs.\,\eqref{eq:anomQ} and \eqref{eq:anomQ1} 
we identify the trace anomaly,     
\beq
\label{eq:trace}
\big(\vartheta_{\mu}^{\mu}\big)_{\rm anom}=
 \frac{R}{2\pi}\big(\partial_{\mu}\phi^{\dagger}\partial^{\mu}\phi-|m|^{2}\phi^{\dagger}\phi\big)
 \,,
 \eeq
while the terms with the $\gamma_{5}$ matrix produce
the axial anomaly  
\beq
\label{acur}
\begin{split}
\big(\partial^{\mu}{\cal A}_{\mu}\big)_{\rm anom}&=-\frac{1}{2\pi}\,\partial_{\mu}
(R\bar\psi\gamma^{\mu}\gamma_{5}\psi)+
i\,\frac{R}{\pi}\,\varepsilon^{\mu\nu}\partial_{\mu}\phi^{\dagger}\partial_{\nu}\phi
\\[1mm]
&=-\frac{g_{0}^{2}}{2\pi}\,\partial^{\mu}{\cal A}_{\mu}
+
i\,\frac{R}{\pi}\,\varepsilon^{\mu\nu}\partial_{\mu}\phi^{\dagger}\partial_{\nu}\phi
\,.
\end{split}
\eeq
In the second line we again use
the relation between the Ricci tensor and the metric, $R=g_{0}^{2}G$
(this is for CP(1)),
to write the fermionic part  of anomaly as  the divergence of the axial current.

The occurrence of the fermionic term $\partial^{\mu}{\cal A}_{\mu}$ in the anomaly \eqref{acur}
is an interesting feature of supersymmetry. The same feature is visible in the general equation 
\eqref{supanom1p}, it explains that we have one and the same coefficient for the fermionic part of the axial anomaly and that in the central charge. Note that a similar phenomenon occurs 
in 4D SYM \cite{grisaru}.
 
From terms linear in $\gamma^{\mu}$ in Eqs.\,\eqref{eq:anomQ} and \eqref{eq:anomQ1} we read off anomalous additions to the central charge density,
 \beq
 \label{Oanom}
 O_{\rm anom}=
 -\frac{1}{2\pi}\,\Big[mg_{0}^{2} D+\frac 1 2\, R\,\bar \psi
(1-\gamma_{5})\psi\Big]=
-\frac{1}{2\pi}\,\Big[mg_{0}^{2} D-i R\psi_{R}^{\dagger}\psi_{L}
\Big]\,.
 \eeq
In the comparison we used the relations
\beq
\gamma^{\mu}\partial_{\mu}=\gamma^{\mu}\gamma_{5}\varepsilon_{\mu\nu}\partial^{\nu}\,,
\qquad
 R\,\partial^{\nu}(\phi^{\dagger}\phi)=g_{0}^{2}\,\partial^{\nu} D\,.
 \eeq
Equation \eqref{Oanom} together with the canonical part $O_{\rm canon}$ from 
Eq.\,\eqref{eq:canO} leads to the expression \eqref{cerfive} for the central charge 
$Z$  quoted in the beginning of this section.

Let us add that, besides supersymmetry used above, there is one more independent 
check of the expression \eqref{Oanom}. Namely, it should 
be renormalization-group invariant. At one loop this is an easy exercise. All one has to do is to calculate two tadpole graphs 
--- one with the fermion loop and another with the boson one.
The tadpole graphs are logarithmically divergent. An appropriate regularization is provided
e.g.\ by the Pauli-Villars scheme.

Omitting simple details of the calculation in the constant background field $\phi$ 
we give here only final results.
The fermion tadpole yields
\beq
R\, \bar \psi\, \frac{1-\gamma_{5}}{2} \,\psi \to -\frac{g_{0}^2}{4\pi}\,m\, \frac{1-\phi^{\dagger}\phi}{1+\phi^{\dagger}\phi}\,\log\frac{|m_{R}|^2}{|m|^2}\,,
\eeq
where $m_{R}$ is the regulator mass.
For the boson tadpole we get
\beq
 mg_{0}^{2}D \to 2m\,\frac{1-\phi^\dagger\,\phi}{(1+\phi^\dagger\,\phi)^3}\,\delta\phi^\dagger\,\delta\phi
\to \frac{g_{0}^2}{4\pi} \,m\,\frac{1-\phi^\dagger\,\phi}{1+\phi^\dagger\,\phi} \,\log\frac{|m_{R}|^2}{|m|^2}\,.
\eeq
The tadpoles cancel each other in the sum.

\section{The Pauli--Villars regularization}
\label{pvreg}
\setcounter{equation}{0}

Strictly speaking, the dimensional regularization used for calculating the superconformal
anomaly is not fully compatible with supersymmetry. Although the problem probably does not
arise at the one-loop level it is better to have an explicitly supersymmetric regularization.
The Pauli--Villars regularization looks  appropriate  for the one-loop considerations.
The problem with it is that for heavy regulator superfields we need to add a superpotential 
to the theory, but, as a rule, this breaks $U(1)$ isometries which we need to preserve in 
the twisted mass theory.

The root of the problem is clearly seen in the framework of  3+1 dimensions
where the theory of one chiral field is anomalous with respect to interaction with 
the U(1) gauge field. This anomaly   produces no problem in reduction to 
the two-dimensional twisted theory at the classical level, but when it comes to regularization
we see a similarity with the four-dimensional case. 

This comparison gives us a hint. To get rid of the gauge anomaly in 3+1 dimensions we should 
add an extra chiral field with the opposite U(1) charge. Let us try the same trick 
in the 1+1 theory adding in the CP(1) model \eqref{cpone} an extra chiral superfield 
with  the mass parameter $m$ of the opposite sign,
\beq
\begin{split}
{\cal L}_{\rm double}&=\int\!{\rm d}^{4}\theta \,\big[K_{m}(\Phi_{1}^{\dagger}{\rm e}^{V}\! \Phi_{1})
+K_{-m}(\Phi_{2}^{\dagger}{\rm e}^{-V}\! \Phi_{2})\big]
\\[1mm]
&=\frac{2}{g_{0}^{2}}\int\!{\rm d}^{4}\theta\,\big[ \log\big(1+\Phi_{1}^{\dagger}{\rm e}^{V}\! \Phi_{1}\big)+\log\big(1+\Phi_{2}^{\dagger}{\rm e}^{-V }\! \Phi_{2}\big)\big]\,.
\end{split}
\eeq
At the classical level we have just two non-communicating CP(1). 
We can add now  the superpotential ${\cal W}(\Phi_{1},\Phi_{2})$ which mixes the $\Phi_{1,2}$ fields,
\beq
\Delta{\cal L}_{\rm double}=\int\!{\rm d}^{2}\theta\,{\cal W}(\Phi_{1},\Phi_{2})+{\rm h.c.}=\frac{2}{g_{0}^{2}}
\,m_{0}\int\!{\rm d}^{2}\theta \,\Phi_{1}\Phi_{2}+{\rm h.c.}\,.
\eeq
This superpotential preserves U(1) symmetry and introduces in addition to the twisted mass 
$m$ a``normal'' mass $m_{0}$ which mixes $\Phi_{1}$ and $\Phi_{2}$.
We are not going to modify the CP(1) model, so we put $m_{0}=0$, but we will add a similar superpotential  for the Pauli--Villars regulators to make them heavy.\footnote{%
\,Note that this is a particular case of the construction of Ref.\  \cite{DHT} which introduces unequal 
number of fields of the opposite U(1) charges. } 

Technically this means that we use the background
field technique for the one-loop calculation
with the Lagrangian which is quadratic in quantum fields  and has the following form:
\beq
\begin{split}
{\cal L}_{\rm reg}^{(2)}=\frac{2}{g_{0}^{2}}&\int\!{\rm d}^{4}\theta\,
\big(1+\Phi_{1}^{\dagger}{\rm e}^{V}\! \Phi_{1}\big)^{-2}
\Big[\delta\Phi_{1}^{\dagger}{\rm e}^{V }\! \delta\Phi_{1}-\frac 1 2
\big(\delta\Phi_{1}^{\dagger}{\rm e}^{V }\! \Phi_{1}\big)^{2}
-\frac 1 2
\big(\Phi_{1}^{\dagger}{\rm e}^{V }\! \delta\Phi_{1}\big)^{2}
\\[1mm]
&+R_{1}^{\dagger}{\rm e}^{V }\! R_{1}-\frac 1 2
\big(R_{1}^{\dagger}{\rm e}^{V }\! \Phi_{1}\big)^{2}
-\frac 1 2
\big(\Phi_{1}^{\dagger}{\rm e}^{V }\! R_{1}\big)^{2}
+\big(1\to 2,\, V\to -V\big)
\Big]
\\[1mm]
&+\Big[\,\frac{2}{g_{0}^{2}}\,M\!\int\!{\rm d}^{2}\theta\, R_{1}R_{2}+{\rm h.c.}\Big]
\,.
\end{split}
\eeq
Here $\delta\Phi_{i}$ are the quantum deviations from the external fields $\Phi_{i}$, and 
$R_{i}$ are corresponding regulator fields. The regulator  fields are quantized abnormally (by anticommutators for bosons and commutators for fermions), so their loops have the opposite sign and regulate the light-field loops.
The cut-off parameter $M$ enters the regulator mass.

Let us start with the one-loop calculation of renormalization of the coupling constant
choosing the background fields in a very simple form,
\beq
\Phi_{1}=F\theta^{2}\,,\qquad \Phi_{2}=0\,,
\eeq
where $F$ is a constant. Then the component form of ${\cal L}_{\rm reg}^{(2)}$  is
\beq
\begin{split}
{\cal L}_{\rm reg}^{(2)}=\frac{2}{g_{0}^{2}}&
\Big[\partial_\mu \phi_{1}^{\dagger}\, \partial^\mu\phi_{1} -(|m|^{2}+2|F|^{2}) \, \phi_{1}^{\dagger}\,\phi_{1} +\partial_\mu \phi_{2}^{\dagger}\, \partial^\mu\phi_{2} -|m|^{2} \, \phi_{2}^{\dagger}\,\phi_{2}
\\[1mm]
&+i\bar \psi_{1} \gamma^{\mu}\partial_{\mu}\psi_{1}
 -\bar \psi_{1}\,\mu\,\psi_{1}+i\bar \psi_{2} \gamma^{\mu}\partial_{\mu}\psi_{2}+\bar \psi_{2}\,\mu\,\psi_{2}
 \\[2mm]
&+ \partial_\mu r_{1}^{\dagger}\, \partial^\mu r_{1} -(m_{R}^{2}+2|F|^{2}) \, { r_{1}^{\dagger}\,r_{1}} +\partial_\mu r_{2}^{\dagger}\, \partial^\mu r_{2} -m_{R}^{2}\, { r_{2}^{\dagger}\,r_{2}}
\\[1mm]
&+i\bar \eta_{1} \gamma^{\mu}\partial_{\mu} \eta_{1}
 -\bar  \eta_{1}\,\mu\, \eta_{1}+i\bar  \eta_{2} \gamma^{\mu}\partial_{\mu} \eta_{2}+\bar  \eta_{2}\,\mu\, \eta_{2} +\big(iM \eta_{1}\gamma^{0} \eta_{2}+ {\rm h.c.}\big)
 \Big]\,,
\end{split}
\eeq
where $\phi_{i}$, $\psi_{i}$ are bosonic and fermionic components of $\delta \Phi_{i}$
and $r_{i}$, $\eta_{i}$ are the same for $R_{i}$\,. We denote $m_{R}$ the regulator mass,
\beq 
m_{R}^{2}=|m|^{2}\!+|M|^{2}\,.
\eeq
 When integrating over quantum fields the boson and fermion loops do not cancel each other only due to
the additional $|F|^{2}$ piece in the bosonic masses 
of the $\phi_{1}$ and $r_{1}$ fields. Thus, integrating out the quantum fields implies 
the following one-loop correction:
\beq
{\cal L}_{\rm one-loop}=-2i|F|^{2}\!\int\!\frac{{\rm d}^{2}k}{(2\pi)^{2}}\left[\frac{1}{k^{2}-|m|^{2}}-
\frac{1}{k^{2}-m_{R}^{2}}\right]=-\frac{|F|^{2}}{2\pi}\log \frac{m_{R}^{2}}{|m|^{2}}\,,
\eeq
where we retain only linear in $|F|^{2}$ terms.
For the chosen background the original Lagrangian  is
\beq
{\cal L}_{0} =\frac{2}{g_{0}^{2}}|F|^{2},
\eeq
so we obtained the coupling constant renormalization,
\beq
\frac{2}{g^{2}}=   \frac{2}{g_{0}^{2}}-\frac{1}{2\pi}\log \frac{m_{R}^{2}}{|m|^{2}}\,,
\eeq
in this particular regularization scheme.  While we use a special background, the reparametrization
invariance allows us to generalize the result to arbitrary backgrounds.

It is simple then to get the dilatation anomaly differentiating over the regulator mass,
\beq
\left(\vartheta^{\mu}_{\mu}\right)_{\rm anom}=-m_{R}\,\frac{\rm d}{{\rm d} m_{R}}\,{\cal L}_{\rm one-loop}= \frac{g_{0}^{2}}{2\pi}\,{\cal L}\,.
\eeq
The result coincides, of course, with Eqs.\,\eqref{eq:trace1} and \eqref{eq:trace}. Supersymmetry relates the dilatation anomaly to other anomalies, including the one in the central charge,  as we discussed in the two previous sections. In Section \ref{sec:supermult} we discussed the supermultiplet of anomalies and its description in the superfield form while in Section \ref{sec:super} we did it starting from the
superconformal anomaly.  Of course, the Pauli--Villars regularization can be used instead of 
dimensional regularization to calculate the
superconformal anomaly and then the central charge anomaly similarly to Section \ref{sec:super}.
We omit presentation of this exercise here. 

\section{Ultraviolet regularization through higher\\ derivatives}
\label{uvreg}
\setcounter{equation}{0}

Our aim  in this Section is  a direct calculation of the anomalous supersymmetry commutator.
We adapt the method of higher derivatives for ultraviolet regularization
following closely an earlier application of the method \cite{SVV} to 
${\cal N}=1$ two-dimensional Landau--Ginzburg models. 

Supersymmetry is explicitly preserved by this regularization, a real advantage of the method,
but Lorentz invariance as well as the reparametrization invariance in the target space are lost.
The advantage of  introducing only spatial derivatives is that the
canonical formalism is essentially unchanged.
The breaking of Lorentz covariance does lead to some ambiguities, to be discussed below.
The requirement of the Lorentz symmetry restoration in the limit of $M\to \infty$ fixes
the ambiguity.

There is one more problem with the method of higher derivatives: while it regularizes the theory,
i.e.\ calculation of amplitudes at any loop order,
it {\em does not} regularize matrix elements of currents at one loop. The problem is well known 
in the case of gauge theories, additional Pauli-Villars regulators are needed to fix one-loop calculations.
As we will show below, one can avoid  explicit introduction of the Pauli-Villars regulators
in the case of the central charge anomalies; this is similar to the consideration in Ref.\cite{SVV}.

It proves sufficient for regularization 
to modify only the bilinear in the superfields $\Phi$, $\Phi^{\dagger}$ part of the CP(1)  Lagrangian.
In terms of the K\"ahler potential this means that 
\beq
\label{eq:K}
K_{\rm reg}=K_{m}+\Delta K\,, \quad K_{m}=
\log\big(1+\Phi^{\dagger} {\rm e}^{V}\Phi\big)\,,\quad
\Delta K=  -\frac{1}{M^{2}}\,\Phi^{\dagger} {\rm e}^{V}\partial_{z}^{2}\Phi\,.
\eeq
Here $\partial_{z}$ is the spatial derivative
and $M$ is the regulator mass to be removed at the very end.
To simplify notations we put $g_{0}^{2}=2$, the one-loop results we are after 
do not contain $g_{0}^{2}$ anyway.
In terms of component fields, one has:
\beq
\label{cponeDel}
\begin{split}
&{\cal L}_{m}=
G\,\Big\{{\cal D}_M \phi^{\dagger}{\cal D}^M \!\phi+i\bar \psi \gamma^{M}\!D_{M}\psi
+\big(F^{\dagger}\!+\frac{i}{2}\,\bar\Gamma\,\psi^{\dagger}\gamma^{0}\psi^{\dagger}\big)
\big(F\!-\frac{i}{2}\,\Gamma\,\psi\gamma^{0}\psi\big)
+\frac{R}{2}(\bar \psi \psi)^{2}\Big\}
\\[2mm]
&\Delta {\cal L}=-\frac{1}{M^{2}}\,\Big\{
{\cal D}_M \phi^{\dagger}\, {\cal D}^M \partial_{z}^{2}\phi 
+i\bar \psi \gamma^{M} {\cal D}_M \partial_{z}^{2}\psi 
+F^{\dagger}\partial_{z}^{2}F\Big\}\,.
\end{split}
\eeq
The expression for ${\cal L}_{m}$ differs from Eq.\,\eqref{cpone} by restoring 
the dependence on the auxiliary field $F$; the expression of this field through others is modified by $\Delta {\cal L}$
in the regularized theory. Let us remind that 
in our notation the mass terms reside 
in the extra components of the covariant derivatives.

The modified equations of motion become
\beq
\label{eq:eom}
\begin{split}
&G\Big[D^{M}{\cal D}_{M}\phi+iR\,{\cal D}_{M}\phi\,\bar \psi \gamma^{M}\psi
-\bar\Gamma\big(F^{\dagger}\!-\frac{i}{4}\,\bar\Gamma\,\psi^{\dagger}\gamma^{0}\psi^{\dagger}\big)
\big(F\!-\frac{i}{2}\,\Gamma\,\psi\gamma^{0}\psi\big)
\\[1mm]
&-\frac{i}{2}\,R\,\psi\gamma^{0}\psi\big(F^{\dagger}\!-\frac{i}{2}\,\bar\Gamma\,\psi^{\dagger}\gamma^{0}\psi^{\dagger}\big)
\Big]-
\frac{1}{ M^{2}}\,\Big[\partial_{z}^{2}{\cal D}^{M}{\cal D}_{M}\phi
+i\bar\Gamma\bar\psi\partial_{z}^{2}\gamma^{M}\!{\cal D}_{M}\psi\Big]
=0\,,
\\[1mm]
&G\Big[i\gamma^{M}\!D_{M}\psi+R\,\psi(\bar\psi\psi)
+i\bar\Gamma\psi^{\dagger}\big(F\!-\frac{i}{2}\,\Gamma\,\psi\gamma^{0}\psi\big)\Big]-
\frac{i}{M^{2}}\,\partial_{z}^{2}\gamma^{M}{\cal D}_{M}\psi=0\,,
\\[1mm]
&G\big(F\!-\frac{i}{2}\,\Gamma\,\psi\gamma^{0}\psi\big)-
\frac{1}{ M^{2}}\,\partial_{z}^{2}F=0\,.
\end{split}
\eeq
From the linearized form of these equations we see that 
in the regularized theory both the bosonic and fermionic propagators acquire
an extra factor $M^{2}/(p_{z}^{2}+M^{2})$. Since 
the vertices are not modified,
the modification of the propagators makes all relevant diagrams convergent.

The supercurrent should also be modified, the original one in Eq.\,\eqref{eq:superJ}
which can be written as 
\beq
J^{\mu}=\sqrt{2}\,G\, {\cal D}_{M}\phi^\dagger \gamma^{M}\gamma^{\mu}\psi
\label{eq:superJ2}
\eeq
is no longer conserved. Let us add to it 
\beq
\label{eq:d1}
\Delta_{1}J^{\mu}=-\frac{\sqrt{2}}{ M^{2}} \,{\cal D}_{M}\phi^\dagger \gamma^{M}
\gamma^{\mu}\partial_{z}^{2}\psi\,,
\eeq
whose time component $\Delta_{1}J^{0}$ follows from $\Delta {\cal L}$ in Eq.\,\eqref{cponeDel}
as the Noether current. Because of the Lorentz invariance breaking  it is still not sufficient 
for current conservation. Indeed, using Eq.\,\eqref{eq:eom} we find
\beq
\partial_{\mu}\big (J^{\mu}+\Delta_{1}J^{\mu}\big)=-\frac{\sqrt{2}}{ M^{2}}\, 
\partial_{z}\big[({\cal D}_{M}
{\cal D}^{M}\!\phi^{\dagger})\!\stackrel{\leftrightarrow}{\partial_z}\!\psi\big]\,.
\eeq
This means that we get the conserved supercurrent 
\beq
J^{\mu}_{\rm reg}=J^{\mu}+\Delta_{1}J^{\mu}+\Delta_{2}J^{\mu}
\eeq
adding to $J^{\mu}$, Eq.\,\eqref{eq:superJ2}, and $\Delta_{1}J^{\mu}$, Eq.\,\eqref{eq:d1} an extra part 
\beq
\Delta_{2}J^{\mu}=\delta^{\mu}_{1}\,\frac{\sqrt{2}}{M^{2}}\,
({\cal D}_{M}{\cal D}^{M}\!\phi^{\dagger})\!\stackrel{\leftrightarrow}{\partial_z}\!\psi\,,
\eeq
which contributes only to the spatial component of $J^{\mu}$.

The construction of the conserved current above is not uniquely defined ---
one can add to $J^{\mu}$ terms of the type $\varepsilon^{\mu\nu}\partial_{\nu} f$ 
which are automatically 
conserved. In other words, one gets a different Noether current moving 
the action of $\partial_{z}$
from $\Phi$ to $\Phi^{\dagger}$ in the expression \eqref{eq:K} for $\Delta K$. While integration by parts
does not affect the theory, it does change the form of the current. This ambiguity is resolved 
by the requirement of Lorentz invariance in our final results. 
We will see that the above choice 
satisfies this condition.

Once the regularized current is constructed, we can find the current of the central charge, $\zeta^{\mu}$,
by the supertransformation,
\beq
\label{eq:zet}
\zeta^{\mu}=\frac{1}{2}\,{\rm Tr}\,\frac{1-\gamma_{5}}{2}\,\big\{J^{\mu}_{\rm reg},\bar Q\big\}\,.
\eeq
Although we performed calculations 
of the central charge anomaly in the generic case, to simplify the presentation we give below 
only the limit
of vanishing twisted mass $m$, and also will work near the origin of the target space, 
$\phi=\phi^{\dagger}=0$. There is no canonical part in the central charge in this limit,
only the anomalous bifermion part.

The anticommutator \eqref{eq:zet} can be calculated using Eq.\,\eqref{eq:a3} from the Appendix.
Although adding higher derivatives changes the canonical quantization, supertransformations of 
all fields stay the same.  What changes is the expression for the auxiliary field $F$. 
Instead of 
Eq.\,\eqref{eq:F}  the last equation in \eqref{eq:eom} should be used. Using also the other equations of motion in Eq.\,\eqref{eq:eom} we arrive at
\beq
\begin{split}
\zeta^{\mu}&=\,-\frac{2}{M^{2}}\,\partial_{z}\Big[\frac{1}{1-\partial_{z}^{2}/M^{2}} \big((\bar \psi \psi) \bar\psi\big)\gamma^{\mu}
\,\frac{1-\gamma_{5}}{2}\!\stackrel{\leftrightarrow}\partial_{z}\!\psi\Big]
\\[1mm]
&+\,\delta^{\mu}_{1}\frac{2}{M^{2}}\,\partial_{\nu}\Big[\frac{1}{1-\partial_{z}^{2}/M^{2}} \big((\bar \psi \psi) \bar\psi\big)
\gamma^{\nu}
\,\frac{1-\gamma_{5}}{2}\!\stackrel{\leftrightarrow}\partial_{z}\!\psi\Big]+\ldots\,,
\end{split}
\eeq
where dots denote terms containing higher powers of the bosonic fields.
Comparing temporal and spatial components of $\zeta^{\mu}$ we see that
\beq
\zeta_{\mu}=-\frac{2}{M^{2}}\,\varepsilon_{\mu\nu}\partial^{\nu}\Big[\frac{1}{1-\partial_{z}^{2}/M^{2}} \big((\bar \psi \psi) \bar\psi\big)\gamma^{0}
\,\frac{1-\gamma_{5}}{2}\!\stackrel{\leftrightarrow}\partial_{z}\!\psi\Big]+\ldots\,.
\eeq
It is now  simple to calculate the fermion tadpole,
\beq
\begin{split}
\zeta_{\mu}&=-\frac{2}{M^{2}}\,\varepsilon_{\mu\nu}\partial^{\nu}\big[\bar \psi(1-\gamma_{5})\psi\big]\int \frac{{\rm d}^{2}p}{(2\pi)^{2}}\,
\frac{p_{z}^{2}}{p^{2}(1+p_{z}^{2}/M^{2})^{2}}+\ldots
\\[1mm]
&=\frac{i}{2\pi}\,\varepsilon_{\mu\nu}\partial^{\nu}\big[\bar \psi(1-\gamma_{5})\psi\big]+\ldots\,.
\end{split}
\eeq
This result is consistent with 
the previous expressions for the anomaly in the limit \mbox{$\phi,\phi^{\dagger}\to 0$}. Its Lorentz covariant form confirms our choice of the regularized current; other choices 
break this.

\section{The curve of  marginal stability}
\label{cms}
\setcounter{equation}{0}

In this section we consider the spectrum of BPS states in CP(1) following the analysis \cite{Dor}.
There is a striking similarity between 
the CP(1) case and  the Seiberg--Witten solution \cite{SW1,SW2}
for ${\cal N}=2$ SQCD in 4D with the SU(2) gauge group and two flavors.
Of particular interest for us is the curve of the marginal stability (CMS)
in the plane of 
the complex mass parameter $m^{2}$ --- a curve where a
restructuring of the BPS states occurs. 
 The  CP(1) model is quite instructive because we deal with elementary functions in this case instead of elliptic integrals in the general case.

The expectation value of the central charge $Z$ over a BPS state can be presented in  CP(1) as
\beq
\label{eq:central}
\big\langle Z \big\rangle_{q,T}= q\, m+T\,m_{D}\,,
\eeq
where $q$ and $T$ are integers corresponding to the Noether and topological charges and \cite{Dor}
\beq
m_{D}=-\frac{i}{2\pi}
\left\{m\log\frac{m+\sqrt{m^2+4\Lambda^2} }
{m- \sqrt{m^2+4\Lambda^2}} -
2\sqrt{m^2+4\Lambda^2}\right\}.
\eeq
Note that the adequate variable is $m^{2}$ rather than $m$.
Indeed, changing the
sign of $m$ is equivalent to the shift 
\beq
\theta \to \theta +2\pi\,.
\label{dopo}
\eeq
Thus, the physical sheet of the Riemann surface is the complex plane of $m^2$,
for $m$ it would be half-plane. In this aspect we differ from Ref. \cite{Dor} where
the full complex plane of $m$ was considered. 

\begin{figure}[h]
 \centerline{\includegraphics[width=4in]{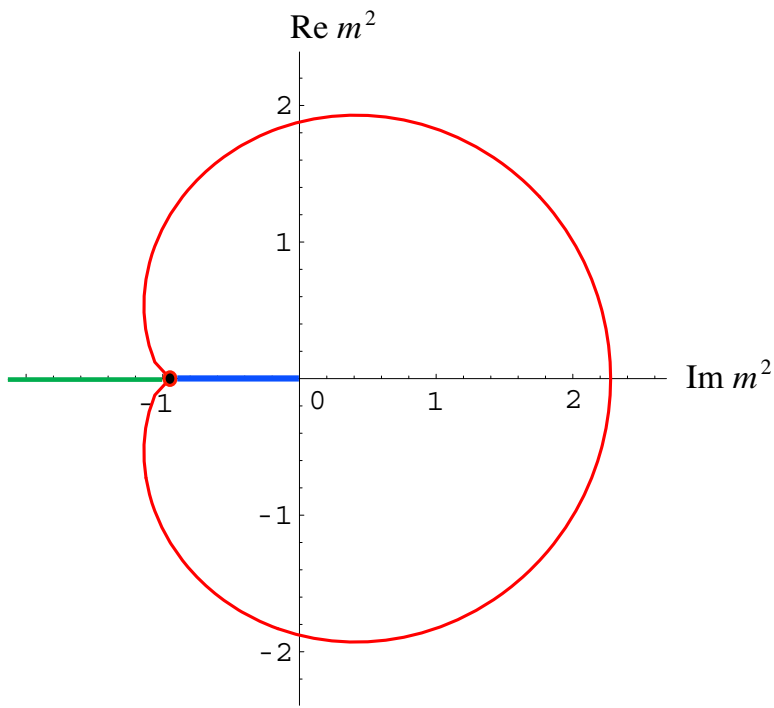}}
 \caption{\small Curve of   marginal stability in CP(1) with twisted mass.
 We set $4\Lambda^2\to 1$. }
 \label{marginal}
 \end{figure}

The complex plane of $m^{2}$
has a cut along the negative 
horizontal axis as it shown in Fig.\,\ref{marginal}. 
When comparing $m$ and $m_{D}$ on the opposite edges of the cut we observe monodromy
around infinity, 
\beq
(m,\, m_{D})\to (-m,\, -m_{D}-m)\,.
\eeq
Correspondingly,
\beq
(q,\, T)\to (-q+T,\, -T)\,.
\eeq

For the BPS state its mass is just $|\langle Z \rangle_{q,T}|$, and the question is  which 
integers $q$ and $T$ correspond to physical stable states at a given value of the parameter
$m^{2}$.
Let us start with the range of large mass, $|m^{2}/4\Lambda^{2}|\gg1$. 
In this range  we are at weak coupling (quasiclassical domain) where
the theory at hand has a rich spectrum of BPS states.
Namely, we have  light ``elementary states" with $T=0$ and
$q=\pm 1$ and  heavy solitons with topological number $T=\pm 1$ and arbitrary integer
value of $q$. 
Each soliton comes with an infinite tower of stable BPS states corresponding to all possible 
values of $q$, similar to dyons in SYM.

On the other hand, at  $|m^{2}/4\Lambda^{2}|\ll 1$ we are at strong coupling.
It is well known \cite{zam,Dor} that in this domain the only BPS states that survive in the spectrum
are $$\{ T= 1,\,\, q =0\}\quad{\rm and }\quad \{ T= -1,\,\, q =1\}$$ (together with their antiparticles, of course).
One of these states becomes massless at \mbox{$m^{2}=-4\Lambda^{2}$}, more precisely,
the $\{ T= 1,\,\, q =0\}$ soliton on the upper side of the cut, and $\{ T= -1,\,\, q =1\}$ on the lower side.
Thus, there is only one singular point $m^{2}=-4\Lambda^{2}$ in the $m^{2}$-plane.
This is different from SU(2) SYM where there are two singular points, $u=\pm \Lambda^{2}$.

Restructuring of the BPS spectrum proves that  the weak coupling domain 
must be separated from strong coupling by a
 curve of  marginal stability. That the CMS exists was shown in
 \cite{Dor}, where it was not explicitly found, however.
 We close this gap here.
 
 The CMS is determined by the condition that the ``electric," $q \,m$, and ``magnetic," $T m_{D}$ parts
 of the exact central charge \eqref{eq:central} have the same phases.
 This is a very simple condition,
 \beq
 {\rm Re}\, \left\{
\log\frac{ 1 +\sqrt{1+4\Lambda^2/m^{2}} }
{ 1 - \sqrt{1+4\Lambda^2/m^2}} -
2\sqrt{1+4\Lambda^2/m^2}
\right\}=0\,.
\label{cms2}
\eeq
The numeric solution to this equation is presented in Fig.~\ref{marginal} where 
$m^2$ is measured in units of  $4\Lambda^2$.
 
 The interval 
 \beq
 {\rm Im}\,  m^{2} =0, \quad {\rm Re}\, m^{2} \in [-1,0]\,
 \eeq
 represents  an analytic solution of Eq.\,(\ref{cms2}).
 However, this interval cannot be reached without crossing the CMS (in fact, it is a part of the cut). If we start, say,
 at large positive $m^{2}$ and travel towards small $m^{2}$ along the real axis, at
 $m^{2}\approx 2.31$ we hit a point where
 the elementary state $\{T=0,\,\,q=1\}$ becomes a marginally bound state of two fundamental solitons
 $\{ T= 1,\,\, q =0\}$ and $\{ T= -1,\,\, q =1\}$.
 At slightly larger $m$ these solitons are bound, at smaller $m$
 attraction changes to
 repulsion, and all towers of states disappear (see \cite{SRVV} for a 
 detailed discussion).
 
 As was mentioned in the beginning of this section there is a direct correspondence between
 CP(1) in 2D and 4D ${\cal N} = 2$ SQCD with SU(2) gauge group and two flavors. 
 The number of variables is, of course, larger in SQCD: besides the moduluar parameter $u$
 we have two mass parameters, $m_{1}$ and $m_{2}$. The correspondence with CP(1) takes 
 place at $u=m_{1}^{2}=m_{2}^{2}$ which is the root of baryonic Higgs branch \cite{Dor}. 
 The massive BPS states in 2D and 4D theories are in one-to-one correspondence upon
 identification $q=n_{e}$ and $T=n_{m}$. 
  
This correspondence is more general. Thus, 2D CP($N\!-\!1$) corresponds to SU($N$) SQCD with 
$N$ flavors in 4D \cite{Dor}. For a generic number of flavors $N_{f}\ge N$ in 4D there is also a 2D counterpart:
a U(1)$_{G}$ gauge theory with $N$ chiral fields with charge $+1$ and $N_{f} -N$ chiral fields 
with charge $-1$ and twisted masses  \cite{DHT}. Moreover, extending the 4D gauge group to SU($N)\times$U(1) allows one to eliminate
the constraint on the matter mass parameters (e.g. for SU(2)$\times$U(1) one can consider
$m_{1}^{2} \neq m_{2}^{2}$) \cite{SY,HT}.  The latter is particularly instructive: the 2D theory 
emerges from the 4D one as a low-energy effective theory on the world sheet of the non-Abelian 
string (flux tube) which is a BPS soliton in the 4D theory.

An interesting question related to the 2D\,--\,4D correspondence is what kind of theory one gets at
the point of singularity which in CP(1) is $m^{2}=-4\Lambda^{2}$.
From the 4D point of view at $u=m_{1}^{2}=m_{2}^{2}=-4\Lambda^{2}$ the quark and monopole vacua coalesce, a phenomenon known as the Argyres--Douglas point \cite{AD} where a nontrivial conformal 
field theory arises. One might suspect that the corresponding 2D theory is also nontrivially conformal.
However, arguments based on the mirror representation indicate against this hypothesis \cite{TY}. 
 
\section{Conclusions}
\label{concl}
\setcounter{equation}{0}

In four-dimensional super-Yang--Mills theory Ferrara and Zumino were the first
to point out \cite{FZ} that the axial current, supercurrent and 
the energy-momentum tensor belonged to a supermultiplet described by a 
{\em hypercurrent} superfield. The superconservation of the hypercurrent
is associated with the superconformal invariance of the classical theory.
At the quantum level this invariance is broken by anomalies which 
also form a supermultiplet \cite{grisaru}. Much later it was realized \cite {DS} that
the anomaly supermultiplet contains also the central charge anomaly.

Two-dimensional CP$(N\!-\!1)$ models are known to be
cousins of four-dimensional super-Yang--Mills, which exhibit, 
in a simplified environment, almost all  interesting phenomena
typical of  non-Abelian gauge theories in four dimensions,
such as asymptotic freedom, instantons, spontaneous breaking of chiral symmetry, etc.
\cite{Wcp,NSVZsigma}.
In spite of the  close parallel existing between  non-Abelian gauge theories in four dimensions
and two-dimensional CP$(N\!-\!1)$ models the issue of the anomaly supermultiplet
and hypercurrent equation 
in the twisted mass CP$(N\!-\!1)$ models has never been addressed in full.
Some aspects were analyzed and important fragments reported in the literature
\cite{Wcp,Dor,LS-03} but, to the best of our knowledge,
the full solution was not presented.
 
We constructed the hypercurrent superfield and the superfield of all anomalies,
including that in the central charge.
Thus,  this question is closed.

As a byproduct, we found the curve of marginal stability in CP(1)
in explicit form.
 
 \section*{Acknowledgments}

We are very grateful to Andrei Losev who participated at early stages of this project,
for informative discussions. We thank 
Sasha Gorsky, David Tong, Misha Voloshin and Alexei Yung for helpful discussions.

This work of M.S. and A.V. was 
supported in part by DOE grant DE-FG02-94ER408.
R.Z.\ is supported by the Swiss National Science Foundation and
         in part by the EU-RTN Programme, Contract No.\ HPEN-CT-2002-00311,
   ``EURIDICE.''

\newpage 
\appendix
 \renewcommand{\theequation}{A.\arabic{equation}}
  \setcounter{equation}{0}  
  \section*{Appendix}  
In this Appendix we collect formulae related to the canonical quantization 
of the CP(1) model.

We use the fields $\phi$, $\phi^{\dagger}$, and $\psi$ as canonical coordinates, then the Lagrangian
\eqref{cponeDel} defines the conjugated momenta,
\beq
\pi_{\phi}=G\partial_{t}\phi^{\dagger}+\Gamma\pi_{\psi}\psi\,,
\qquad \pi_{\phi^{\dagger}}=G\partial_{t}\phi\,,
\qquad \pi_{\psi}=iG\bar \psi\gamma^{0}\,.
\eeq
Note the asymmetry between $\phi$ and $\phi^{\dagger}$, and also 
between $\psi$ and $\bar\psi$.
The canonical commutation relations determine the equal-time commutators 
for the fields (and their  time derivatives),
\beq
\begin{split}
&\big[\partial_{t} \phi^{\dagger}(t,z),\phi(t,z')\big]=-iG^{-1}\delta(z-z')\,,\qquad
\big[\partial_{t} \phi^{\dagger}(t,z),\psi(t,z')\big]=iG^{-1}\Gamma\psi\,\delta(z-z')\,,
\\[2mm]
&\big[\partial_{t} \phi(t,z),\phi^{\dagger}(t,z')\big]=-iG^{-1}\delta(z-z')\,,\qquad
\big[\partial_{t} \phi(t,z),\bar\psi(t,z')\big]=-iG^{-1}\bar \Gamma\,\bar\psi\,\delta(z-z')\,,
\\[2mm]
&\big[\partial_{t} \phi(t,z),\partial_{t} \phi^{\dagger}(t,z')\big]=-iG^{-1}\big[\Gamma\partial_{t} \phi
-\bar\Gamma\partial_{t} \phi^{\dagger}+iR\,\bar\psi \gamma^{0}\psi\big]
\delta(z-z')\,, 
\\[2mm]
&\big\{\psi_{\alpha}(t,z),\bar\psi_{\beta}(t,z')\big\}=G^{-1}(\gamma^{0})_{\alpha\beta}\delta(z-z')\,.
\end{split}
\eeq
All other (anti)commutators vanish. 

Using the expression \eqref{eq:superJ} for the supercharges 
we can verify then that the canonical commutators reproduce the SUSY transformations,
\begin{equation}
\begin{split}
&[\phi\,,\bar Q_\beta] = 0\,, \qquad [\phi^\dagger, \bar Q_\beta]   = i  \sqrt{2}\,\bar \psi_\beta\,,\qquad
\{\psi_\alpha\,,\bar Q_\beta\} = \sqrt{2}\,(\slsh{\partial}-i\mu^{\dagger})_{\alpha\beta} \phi \,,
\\[2mm]
&\{\bar\psi_\alpha\,,\bar Q_\beta\} = \sqrt{2}\,\bar F\, (\gamma^{0})_{\alpha\beta}\,,\qquad
[\phi\,, Q_\beta]=i\sqrt{2}\,\psi_{\beta}\,,\qquad [\phi^{\dagger}, Q_\beta]=0\,,
\\[2mm]
&\{\psi_\alpha\,, Q_\beta\} = \sqrt{2}\,F\, (\gamma^{0})_{\alpha\beta}\,,
\qquad
\{Q_\beta,\bar\psi_\alpha\} = \sqrt{2}\,(\slsh{\partial}+i\mu^{\dagger})_{\beta\alpha} \phi \,.
\end{split}
\label{eq:a3}
\end{equation}
Here
\beq
\label{eq:F}
F=\frac{i}{2}\,\Gamma\psi\gamma^{0}\psi\,,\qquad \bar F=F^{\dagger}
\eeq
are the upper components of superfields $\Phi$ and $\Phi^{\dagger}$.

Note that from Eq.\,\eqref{eq:a3} it follows that
\beq
\{G\psi_\alpha\,, Q_\beta\} =0\,,\qquad \{G\bar\psi_\alpha\,,\bar Q_\beta\} =0\,.
\label{eq:a5}
\eeq
We used these relations  to determine the anticommutators of the the currents and 
supercharges.

\end{document}